\documentclass[12pt]{article}
\usepackage{epsfig}

\usepackage{axodraw}

\newcommand{\mysection}{\setcounter{equation}{0}\section}

\def\beq{\begin{equation}}
\def\eeq{\end{equation}}
\def\beqa{\begin{eqnarray}}
\def\eeqa{\end{eqnarray}}

\begin{document}

\begin{center}
{\Large \bf Soft and collinear enhancements to top quark and Higgs cross 
sections}\footnote{Presented at Cracow Epiphany Conference on LHC Physics, 
January 2008}
\end{center}
\vspace{2mm}
\begin{center}
{\large Nikolaos Kidonakis}\\
\vspace{2mm}
{\it Kennesaw State University, Physics \#1202\\
1000 Chastain Rd., Kennesaw, GA 30144-5591}
\end{center}

\begin{abstract}
I present calculations of soft and collinear corrections to the 
cross sections for single top quark production, and for Higgs production via 
$b{\bar b} \rightarrow H$ at the Tevatron and the LHC. 
I show that the corrections provide significant enhancements to the cross 
sections in both cases. For single top production the soft gluon corrections 
dominate the cross section, particularly in the $s$ channel and in 
$tW$ production.
For Higgs production it is shown that purely collinear
terms have to be included as well to provide an accurate calculation.
\end{abstract}
  
\thispagestyle{empty} \newpage \setcounter{page}{2}

\mysection{Introduction}

The top quark and the Higgs boson are elementary particles 
that are the epicenter of a lot of theoretical and experimental study 
due to their significance to electroweak theory and QCD. 
The importance of these particles to the Standard Model  
requires that we understand their cross sections at hadron colliders 
with the highest achievable accuracy.

The top quark is the heaviest known elementary 
particle and was discovered in proton-antiproton collisions at the 
Fermilab Tevatron by the CDF and D0 experiments in 1995 \cite{CDFtt,D0tt}. 
The discovery of the top quark was through the production of top-antitop 
pairs. 
Since then there has been a continuing effort (for a review see 
\cite{WW,Kehoe}) to understand the 
properties of the top and refine the measurement of the value of its mass.
The $t{\bar t}$ cross section has been calculated to high accuracy by 
including next-to-next-to-leading-order (NNLO) threshold corrections 
\cite{NKRVtop}, which dominate the cross section. The experimental 
values for the cross section \cite{CDFcs,D0cs} are in good agreement with the 
theoretical prediction \cite{NKRVtop}.

The top quark can also be produced in single top production 
through $t$-channel processes  
involving the exchange of a space-like $W$ boson, $s$-channel
processes involving the exchange of a time-like $W$ boson, 
and $tW$ production processes involving the production of a $W$ boson 
in association with the top quark.
However the cross section is smaller than for $t{\bar t}$ production  
and the signal is complicated by relatively large backgrounds. 
Nevertheless, there is now recent evidence of single top production 
\cite{D0t,CDFt}.
The interest in this process is due to the fact that it allows 
a measurement of the $V_{tb}$ CKM matrix element and other electroweak 
properties of the top quark, and it may play a role in the discovery of new 
physics.
 
The search for the Higgs boson \cite{Higgs} is currently the most important 
goal at the Tevatron and the LHC colliders \cite{HiggsWG}. In the Standard 
Model it is expected that the dominant production channel at either collider 
will be $gg \rightarrow H$. However,
the channel $b{\bar b}\rightarrow H$ can also be important
in the Minimal Supersymmetric Standard Model at high $\tan\beta$, 
the ratio of the vacuum expectation values for the two Higgs doublets.

Since both the top quark and the Higgs boson are very massive, their 
production cross sections receive large corrections from 
soft and collinear gluon corrections, which can dominate the cross section 
near threshold.
These threshold corrections arise from incomplete cancellations of infrared 
divergences between virtual diagrams 
and real diagrams with soft (low-energy) gluons.
The corrections exponentiate and can thus be resummed. 

For single-top production we have processes of the form 
$p_1 +p_2 \rightarrow p_3+p_4$, and we  
define $s=(p_1+p_2)^2$, $t=(p_1-p_3)^2$, $u=(p_2-p_3)^2$
and $s_4=s+t+u-m_3^2-m_4^2$. At threshold $s_4 \rightarrow 0$. 
The soft corrections take the form
$[\ln^k(s_4/m_t^2)/s_4]_+$, with $m_t$ the top quark mass and 
$k \le 2n-1$ for the ${\cal O} (\alpha_s^n)$ corrections \cite{NKsingletop}.
Near threshold these corrections are dominant and provide excellent 
approximations to the full cross section. 
 
For Higgs production, we define $z=m_H^2/s$, where $m_H$ is the Higgs mass, 
and $z \rightarrow 1$ at threshold. 
The soft corrections now take the form
$[\ln^k(1-z)/(1-z)]_+$ \cite{Ravi,NKHiggs}. 
For $b{\bar b} \rightarrow H$, in addition to the soft corrections, we also 
have to include purely collinear corrections of the form $\ln^k(1-z)$ in order 
to achieve a good approximation \cite{NKHiggs}.
The $n$-th order corrections in the partonic cross section for 
$b{\bar b}\rightarrow H$ can be written as
\beq
{\hat \sigma}^{(n)}(z)=V^{(n)} \, \delta(1-z)+\sum_{k=0}^{2n-1} S_k^{(n)}
\left[\frac{\ln^k(1-z)}{1-z}\right]_+ +\sum_{k=0}^{2n-1} C_k^{(n)}
\ln^k(1-z) \, ,
\nonumber
\eeq
with a similar expression holding for single-top production.
The hadronic cross section, $\sigma$, is calculated by integrating the product 
of the partonic cross section, ${\hat \sigma}$, and parton distribution 
functions, $\phi$, over the momenta fractions of the protons and/or 
antiprotons carried by the partons in the hard scattering, 
at factorization scale $\mu_F$ and renormalization scale $\mu_R$:
\beq
\sigma=\sum_f \int  dx_1 dx_2 \, \phi_{f_1/p}(x_1,\mu_F)\,
\phi_{f_2/{\bar p}}(x_2,\mu_F)\, {\hat \sigma}(z,\mu_F,\mu_R,\alpha_s) \, . 
\eeq

\mysection{Resummed cross section and NNNLO expansions}

The resummation of soft and collinear 
gluon corrections is performed in moment space and 
it follows from the factorization properties of the cross section. 
For the process $b {\bar b} \rightarrow H$ we can write 
the resummed cross section as \cite{NKHiggs,NKuni,NKNNNLO}
\beqa
{\hat{\sigma}}^{res}(N) &=&
\exp\left[2 E_q(N)\right] \; \exp\left[2 E^{coll}_q(N)\right]
\exp \left[4 \int_{\mu_F}^{m_H} \frac{d\mu}{\mu}
\gamma_{q/q}(N,\mu)\right] \;
\nonumber\\ && \times \,
H(\mu_R) \; S \left(\frac{m_H}{\tilde N}\right) \;
\exp \left[\int_{m_H}^{\frac{m_H}{\tilde N}}
\frac{d\mu}{\mu} 2 {\rm Re}\Gamma_S(\mu)\right] 
\label{resHS}
\eeqa
where
\beqa
E_{q}(N)&=&-C_F \int^1_0 dz \frac{z^{N-1}-1}{1-z}\;
\left \{\int^1_{(1-z)^2} \frac{d\lambda}{\lambda}
\frac{\alpha_s(\lambda m_H^2)}{\pi}
+\frac{\alpha_s((1-z)^2 m_H^2)}{\pi}\right\}
\nonumber \\ && 
{}+{\cal O}(\alpha_s^2) 
\label{Eexp}
\eeqa
and a somewhat similar expression holds for single top production 
\cite{NKsingletop}.
Here $C_F=(N_c^2-1)/(2N_c)$ with $N_c$ the number of colors.
The purely collinear logarithms are resummed in the second exponent, 
$E_q^{coll}$,  
which has a form similar to Eq. (2.2) with the substitution 
$-(z^{N-1}-1)/(1-z) \rightarrow z^{N-1}$.  
The quark anomalous dimensions are denoted by $\gamma_{q/q}$,  
$H$ is the hard-scattering factor,  
$\Gamma_S$ is the soft anomalous dimension which describes the evolution 
of the soft function $S$, and $\tilde N= N e^{\gamma_E}$ with $\gamma_E$ 
the Euler constant. 
Inverting Eq. (\ref{resHS}) back to momentum space, 
we now provide next-to-next-to-next-to-leading-order (NNNLO) 
expansions of the resummed cross section.

At next-to-leading order (NLO) the soft and collinear gluon corrections are
\beqa
{\hat\sigma}^{(1)}
&=&F^B
\frac{\alpha_s(\mu_R^2)}{\pi} \left\{
c_{3} \left[\frac{\ln(1-z)}{1-z}\right]_+
+c_{2} \left[\frac{1}{1-z}\right]_+
+c_{1} \, \delta(1-z)\right.
\nonumber \\ && \hspace{20mm} \left. 
{}+ c_3^c \ln(1-z) +c_2^c\right\}
\nonumber
\label{NLO}
\eeqa
where $F_B$ is the Born term. The leading coefficients for 
$b{\bar b} \rightarrow H$ are  $c_3=4C_F$, 
$c_3^c=-4C_F$ and the subleading coefficients are given in 
\cite{NKHiggs}. Expressions for all coefficients in the single top processes 
are given in \cite{NKsingletop}. 

At NNLO the soft and collinear gluon corrections are
\beq
{\hat\sigma}^{(2)}
=F^B \frac{\alpha_s^2(\mu_R^2)}{\pi^2}
\left\{\frac{1}{2} c_3^2
\left[\frac{\ln^3(1-z)}{1-z}\right]_+
+\cdots
+\frac{1}{2} c_3 c_3^c \ln^3(1-z)
+\cdots
\right\}
\label{NNLO}
\nonumber
\eeq
where we show explicitly only the leading logarithms.

Finally, the NNNLO soft and collinear gluon corrections
are
\beq
{\hat\sigma}^{(3)}
=F^B \frac{\alpha_s^3(\mu_R^2)}{\pi^3}
\left\{\frac{1}{8} c_3^3
\left[\frac{\ln^5(1-z)}{1-z}\right]_+
+\cdots
+\frac{1}{8} c_3^2 c_3^c \ln^5(1-z) 
+\cdots \right\}
\nonumber
\label{NNNLO}
\eeq
where again we only show the leading logarithms. Further 
details are provided in Refs. \cite{NKsingletop,NKHiggs}.

\mysection{Single top quark production}

\begin{figure}
\begin{center}
\begin{picture}(80,120)(0,0)
\Line(0,75)(40,70)
\Line(40,70)(80,75)
\Line(0,25)(40,30)
\Line(40,30)(80,25)
\Photon(40,70)(40,30){3}{5}
\Text(0,15)[c]{$b$}\Text(0,85)[c]{$q$ (${\bar q}$)}
\Text(55,50)[c]{$W$}
\Text(80,15)[c]{$t$}\Text(80,85)[c]{$q'$ (${\bar q}'$)}
\Text(40,5)[c]{(a)}
\end{picture}
\hspace{10mm}
\begin{picture}(80,120)(0,0)
\Line(0,75)(20,50)
\Line(0,25)(20,50)
\Line(60,50)(80,75)
\Line(60,50)(80,25)
\Photon(20,50)(60,50){3}{5}
\Text(0,15)[c]{${\bar q'}$}\Text(0,85)[c]{$q$}
\Text(40,65)[c]{$W$}
\Text(80,15)[c]{$t$}\Text(80,85)[c]{${\bar b}$}
\Text(40,5)[c]{(b)}
\end{picture}
\hspace{10mm}
\begin{picture}(80,120)(0,0)
\Line(0,75)(20,50)
\Gluon(0,25)(20,50){3}{5}
\Line(20,50)(60,50)
\Photon(60,50)(80,25){3}{5}
\Line(60,50)(80,75)
\Text(0,15)[c]{$g$}\Text(0,85)[c]{$b$}
\Text(40,65)[c]{$b$}
\Text(80,15)[c]{$W$}\Text(80,85)[c]{$t$}
\Text(40,5)[c]{(c)}
\end{picture}
\end{center}
\vspace{-5mm}
\caption{Leading-order $t$-channel (a), $s$-channel (b), and
associated $tW$ production (c) diagrams for
single top quark production.}
\end{figure}
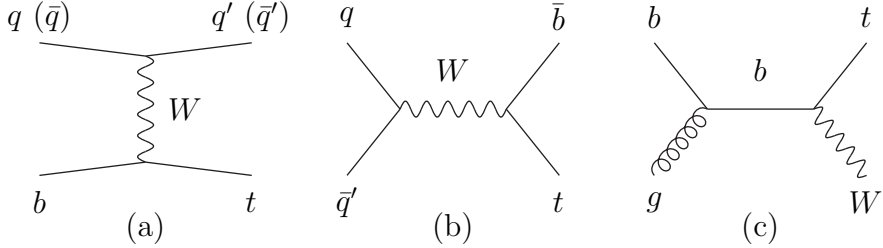

The leading-order (LO) diagrams for the production of single top quarks 
in the $t$-channel, $qb \rightarrow q' t$ and ${\bar q} b \rightarrow 
{\bar q}' t$; $s$-channel, $q{\bar q}' \rightarrow {\bar b} t$; 
and via associated $tW$ production, $bg \rightarrow tW^-$, are shown in 
Fig. 1.

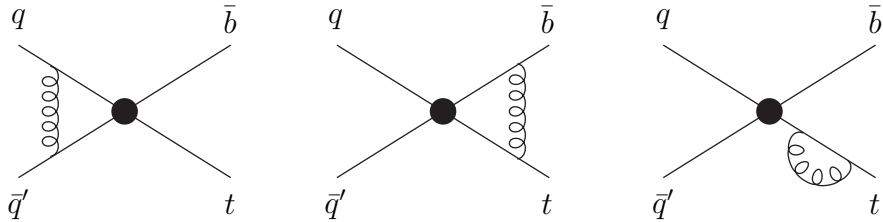
\begin{figure}
\begin{center}
\begin{picture}(240,100)(0,0)
\Line(0,75)(40,50)
\Line(0,25)(40,50)
\Line(40,50)(80,75)
\Line(40,50)(80,25)
\Gluon(12,67)(12,33){3}{5}
\Vertex(40,50){5}
\Text(0,15)[c]{${\bar q'}$}\Text(0,85)[c]{$q$}
\Text(80,15)[c]{$t$}\Text(80,85)[c]{${\bar b}$}
\Line(120,75)(160,50)
\Line(120,25)(160,50)
\Line(160,50)(200,75)
\Line(160,50)(200,25)
\Gluon(188,68)(188,32){3}{5}
\Vertex(160,50){5}
\Text(120,15)[c]{${\bar q'}$}\Text(120,85)[c]{$q$}
\Text(200,15)[c]{$t$}\Text(200,85)[c]{${\bar b}$}
\end{picture}
\begin{picture}(80,120)(0,0)
\Line(0,75)(40,50)
\Line(0,25)(40,50)
\Line(40,50)(80,75)
\Line(40,50)(80,25)
\Vertex(40,50){5}
\GlueArc(60,35)(10,137,340){3}{4}
\Text(0,15)[c]{${\bar q'}$}\Text(0,85)[c]{$q$}
\Text(80,15)[c]{$t$}\Text(80,85)[c]{${\bar b}$}
\end{picture}
\end{center}
\vspace{-5mm}
\caption{One-loop eikonal corrections to the soft function
for the $s$-channel diagram in single top quark production.}
\end{figure}

Next-to-leading logarithm (NLL) resummation requires that we calculate 
one-loop 
corrections in the eikonal approximation to these diagrams. 
The diagrams for these corrections in the $s$ channel are shown in 
Fig. 2. Similar diagrams are used for the $t$ channel and  
$tW$ production \cite{NKsingletop}. 

We now turn our attention to a numerical study of the cross sections for 
single-top production at the Tevatron and the LHC \cite{NKsingletop}.  
We use the MRST 2004 parton densities \cite{MRST2004} in our results.

\subsection{Single top production at the Tevatron} 

We provide below results for single top production at the Tevatron 
with $\sqrt{S}=1.96$ TeV, 
and note that the cross section for single antitop production is identical
to that for the top. 

We begin with the $t$ channel.
The soft-gluon corrections in this channel are relatively small.
To find the best estimate for the cross section we match to the
exact NLO cross section, i.e. we add the soft-gluon 
corrections through NNNLO at NLL accuracy \cite{NKsingletop} 
to the exact NLO cross section \cite{bwhl}.
The matched cross section for a top quark mass of 
170 GeV is
\beq
\sigma^{t-{\rm channel}}(m_t=170 \,{\rm GeV})=1.17^{+0.02}_{-0.01}
\pm 0.06 \; {\rm pb}
\nonumber
\eeq
where the first uncertainty is due to the scale dependence,
derived by varying the scale between $m_t/2$ and $2m_t$,
and the second is due to the parton density uncertainties.
The corresponding quantity for a top quark mass of 175 GeV is 
$1.08^{+0.02}_{-0.01}\pm 0.06$ pb.

\begin{figure}
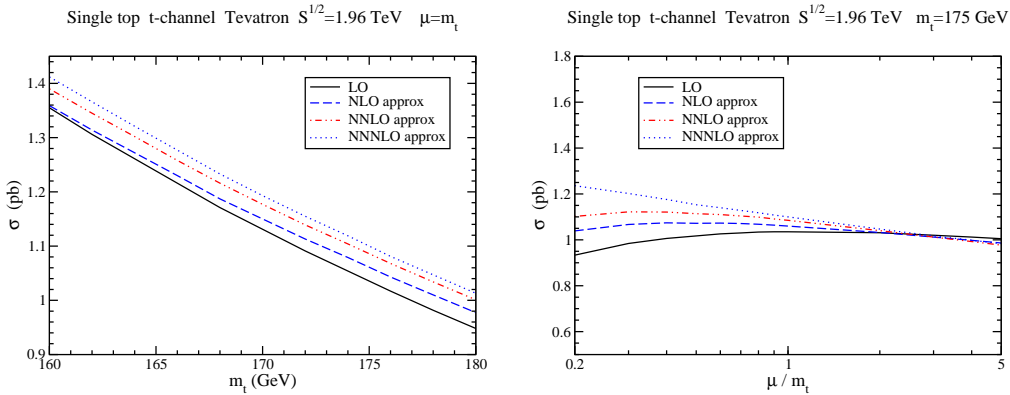

\begin{center}
\epsfig{file=tchtevmtplotactphpol.eps,
       height=0.38\textwidth,angle=0} 
\hspace{3mm}
\epsfig{file=tchtevmu175plotdeltaactphpol.eps,
       height=0.38\textwidth,angle=0}
\caption{The cross section for single top production in the $t$ channel 
at the Tevatron.}
\end{center}
\end{figure}

Figure 3 shows the results for the cross section in the $t$ channel.
On the left we show the LO cross section as well
as the cross sections with the NLO, NNLO, and NNNLO soft-gluon corrections
included versus the top quark mass $m_t$, with the factorization and
renormalization scales set equal to $m_t$.
On the right we plot the scale dependence of the cross section with 
$m_t=175$ GeV.
Here we set the factorization scale equal to the
renormalization scale and vary this scale $\mu$ over a large range. 

We continue with the $s$ channel. The soft-gluon corrections 
approximate the full QCD corrections very well. The corrections are 
relatively large for this channel, in stark contrast with the results 
we found in the $t$ channel.
The matched cross section (exact NLO + soft gluon corrections through NNNLO)
is 
\beq
\sigma^{s-{\rm channel}}(m_t=170 \, {\rm GeV})=0.56 \pm 0.02
\pm 0.01 \; {\rm pb}
\nonumber \\
\eeq
while for $m_t=175$ GeV it is $0.49 \pm 0.02 \pm 0.01$ pb.

\begin{figure}
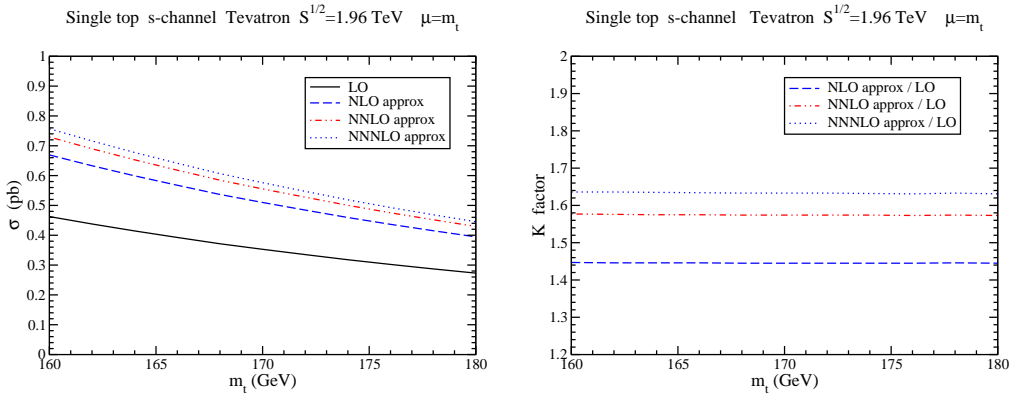

\begin{center}
\epsfig{file=schtevmtplotactphpol.eps,
       height=0.38\textwidth,angle=0}
\hspace{3mm}
\epsfig{file=Kschtevmtplotactphpol.eps,
       height=0.38\textwidth,angle=0}
\caption{The cross section (left) and $K$ factors (right) for single top 
production in the $s$ channel at the Tevatron.}
\end{center}
\end{figure}

In Fig. 4 we plot the cross section
for single top quark production at the Tevatron 
in the $s$ channel as a function of $m_t$ setting the scales to $\mu=m_t$.
On the left we plot the LO cross section
and the approximate NLO, NNLO, and NNNLO cross sections at NLL accuracy.
The $K$ factors are quite large and are shown on the right.

The cross section for single top production at the Tevatron via the $tW$ 
channel is numerically the smallest. 
The approximate NNNLO cross section is 
\beq
\sigma^{tW}(m_t=170 \, {\rm GeV})=0.15 \pm 0.02 \, 
\pm 0.03 \; {\rm pb}
\nonumber \\
\eeq
and it is $0.13 \pm 0.02 \pm 0.02$ pb for $m_t=175$ GeV.

\subsection{Single top production at the LHC} 

We now show results for single top production at the LHC with  
$\sqrt{S}=14$ TeV.
At the LHC the top and antitop cross sections are different in the $t$ and 
$s$ channels.

In the $t$ channel the threshold corrections are not a good approximation of 
the full QCD corrections.
The exact NLO cross section is
$152 \pm 5 \pm 3$ pb for $m_t=170$ GeV, and $146 \pm 4 \pm 3$ pb for 
$m_t=175$ GeV.
For antitop production in the  $t$ channel the exact NLO cross section is
$93 \pm 3 \pm 2$ pb for $m_t=170$ GeV, and $89 \pm 3 \pm 2$ pb for 
$m_t=175$ GeV.

For single top production at the LHC through the $s$ channel the 
matched cross section, i.e. exact NLO plus soft gluon corrections through 
NNNLO, is 
\beq
\sigma^{s-{\rm channel}}_{\rm top}(m_t=170\, {\rm GeV})=
8.0^{+0.6}_{-0.5} \pm 0.1 \; {\rm pb}
\nonumber \\
\eeq
and the corresponding cross section for $m_t=175$ GeV is 
$7.2^{+0.6}_{-0.5} \pm 0.1$ pb.

For single antitop production at the LHC in the $s$ channel
the matched cross section is 
\beq
\sigma^{s-{\rm channel}}_{\rm antitop}(m_t=170 \, {\rm GeV})=
4.5 \pm 0.1 \pm 0.1 \; {\rm pb}
\nonumber \\
\eeq
and it is 
$4.0 \pm 0.1 \pm 0.1$ pb for $m_t=175$ GeV.

\begin{figure}
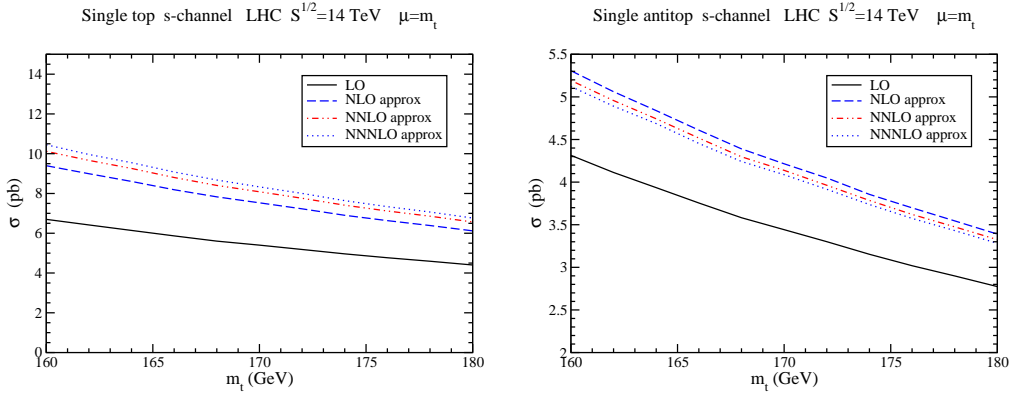

\begin{center}
\epsfig{file=schlhcmtplotactphpol.eps,
       height=0.38\textwidth,angle=0}
\hspace{3mm}
\epsfig{file=aschlhcmtplotactphpol.eps,
       height=0.38\textwidth,angle=0}
\caption{The cross section for single top (left) and single antitop (right) 
production in the $s$ channel at the LHC.}
\end{center}
\end{figure}

In Fig. 5 we plot the cross section
for single top quark (left) and single antitop quark (right) 
production at the LHC in the $s$ channel as a function of $m_t$ 
setting the scales to $\mu=m_t$.

For the $tW$ channel, the matched cross section (exact NLO \cite{Zhu} 
plus NNNLO soft corrections) is 
\beq
\sigma^{tW}(m_t=170 \, {\rm GeV})=44 \pm 5 \pm 1  
\; {\rm pb}
\nonumber \\
\eeq
and the result for $m_t=175$ GeV is  $41 \pm 4 \pm 1$ pb.  
The cross section for anti-top production in this channel is identical to that 
for top.

\begin{figure}
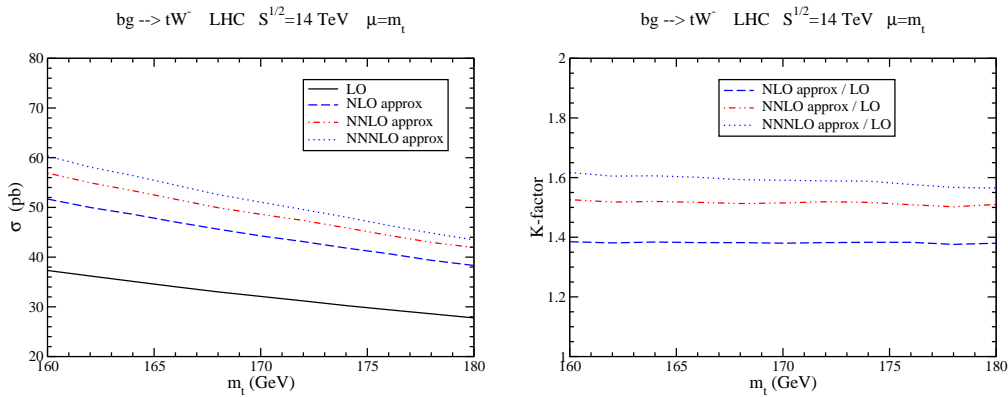

\begin{center}
\epsfig{file=bglhcmtplotactphpol.eps,
       height=0.38\textwidth,angle=0}
\hspace{3mm}
\epsfig{file=Kbglhcmtplotactphpol.eps,
       height=0.38\textwidth,angle=0}
\caption{The cross section (left) and $K$ factors (right)
for $tW$ production at the LHC.}
\end{center}
\end{figure}

In Fig. 6 we plot the cross section (left) and $K$ factors (right)
for $tW$ production at the LHC as functions of $m_t$, setting the scales 
to $\mu=m_t$.
The higher-order corrections are quite significant as shown by the large 
$K$ factors.

\mysection{Higgs production via $b {\bar b} \rightarrow H$}

The process $b{\bar b} \rightarrow H$ has a very simple color structure and 
kinematics, and is very similar to the Drell-Yan process.
The QCD corrections are fully known to NNLO \cite{HKbb}.
We can can calculate all soft corrections fully to NNNLO \cite{Ravi,NKHiggs}.  
However, it is found that the soft-gluon approximation is inadequate:  
purely collinear terms must be included to provide a good approximation.
The collinear terms can be resummed at LL accuracy, and good approximations
at NLL and NNLL accuracy were provided in \cite{NKHiggs}.
Below we provide numerical results for the cross section at the Tevatron and 
the LHC using the MRST 2006 parton densities \cite{MRST2006}.

\begin{figure}
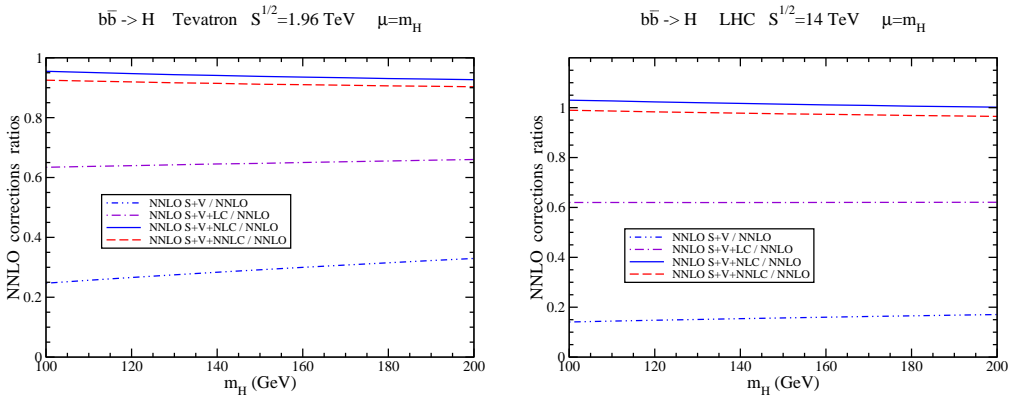

\begin{center}
\epsfig{file=Kbbhtevnnloplotactphpol.eps, 
        height=0.38\textwidth,angle=0}
\hspace{3mm}
\epsfig{file=Kbbhlhcnnloplotactphpol.eps, 
        height=0.38\textwidth,angle=0}
\caption{The NNLO ratios for $b {\bar b} \rightarrow H$ at the Tevatron (left)
and the LHC (right). Here $\mu=\mu_F=\mu_R=m_H$.} 
\end{center}
\end{figure}

In Fig. 7 we show the contribution of various terms
to the complete NNLO corrections for Higgs production via
$b {\bar b} \rightarrow H$ with $\mu=m_H$ at the Tevatron (left-hand side)
and the LHC (right-hand side).
In this figure NNLO denotes the ${\cal O}(\alpha_s^2)$ corrections only, 
without the LO term and NLO corrections.
The curve marked NNLO S+V / NNLO denotes the percentage contribution of the
NNLO soft plus virtual (S+V) corrections to the total NNLO corrections. 
We see that both at the Tevatron and the LHC this contribution is small. 
Inclusion of the leading collinear (LC) logarithms accounts for 
about 60\% of the total NNLO corrections at both the Tevatron and the LHC.
Including the next-to-leading collinear (NLC) logarithms vastly improves
the approximation. 
The difference between the S+V+LC and the S+V+NLC curves is around 30\%
at the Tevatron and  40\% at the LHC; thus the NLC terms are of great
importance in achieving a good approximation.
We also plot a curve (S+V+NNLC) that in addition includes the 
next-to-next-to-leading collinear (NNLC) terms. 
The NNLC terms alone do not make a large
contribution, and the S+V+NNLC results approximate the exact NNLO
corrections very well, especially at the LHC.

\begin{figure}
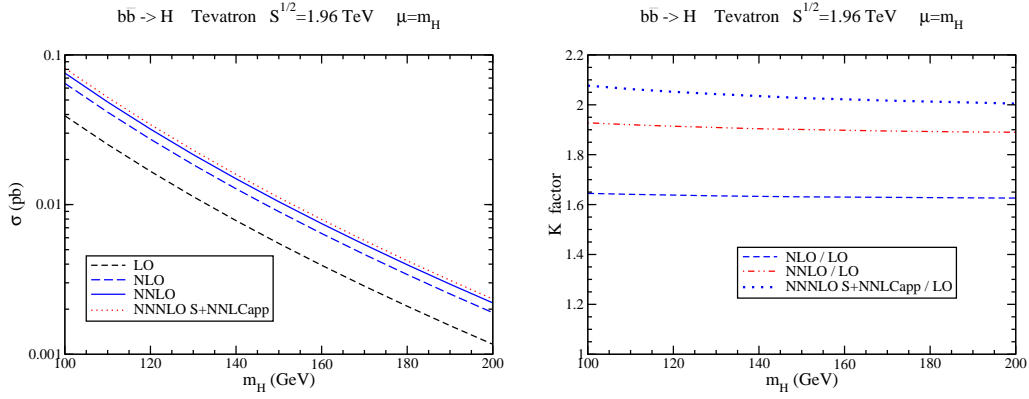

\begin{center}
\epsfig{file=bbhtevplotactphpol.eps, 
        height=0.38\textwidth,angle=0}
\hspace{3mm}
\epsfig{file=Kbbhtevplotactphpol.eps, 
        height=0.38\textwidth,angle=0}
\caption{The cross section (left) and $K$ factors (right) 
for $b {\bar b} \rightarrow H$ at the Tevatron.
Here $\mu=\mu_F=\mu_R=m_H$.}
\end{center}
\end{figure}

In Fig. 8 we plot the cross sections (left) and $K$ factors (right) 
for $b {\bar b} \rightarrow H$ at the Tevatron with $\mu=m_H$. 
The complete NLO corrections increase the LO result by over
60\%. The NNLO corrections futher increase the cross section by roughly an
additional 30\%. Finally, we include the soft and 
purely collinear corrections at NNNLO. The soft corrections are
complete and the purely collinear terms are approximate.
Our study of the contributions of the soft and collinear terms at NNLO  
gives us confidence that the NNNLO S+NNLCapp curves
provide a good approximation of the complete NNNLO cross section.
The NNNLO S+NNLCapp corrections provide an additional 10\% to 
15\% increase to the cross section.

\begin{figure}
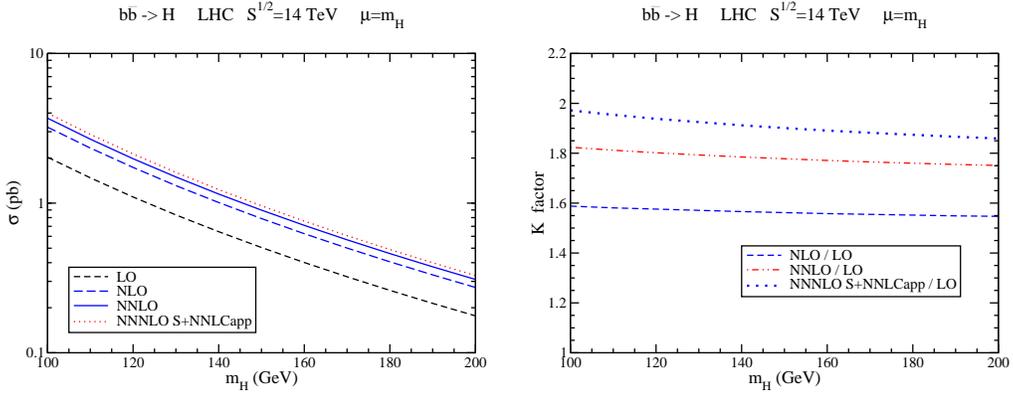

\begin{center}
\epsfig{file=bbhlhcplotactphpol.eps, 
        height=0.38\textwidth,angle=0}
\hspace{3mm}
\epsfig{file=Kbbhlhcplotactphpol.eps, 
        height=0.38\textwidth,angle=0}
\caption{The cross section (left) and $K$ factors (right) 
for $b {\bar b} \rightarrow H$ at the LHC.
Here $\mu=\mu_F=\mu_R=m_H$.}
\end{center}
\end{figure}

In Fig. 9 we plot the cross sections (left) and $K$ factors (right) 
for $b {\bar b} \rightarrow H$ at the LHC for $\mu=m_H$. We see that the 
$K$ factors here are quite similar (slightly smaller) to those 
for the Tevatron.

Finally, we note that the parton distribution function uncertainties 
for this process are non-negligible and can be
of the same order of magnitude as the scale uncertainty, especially for
large Higgs masses at the Tevatron.

\mysection{Conclusion}

The soft and collinear corrections make important contributions 
in the cross sections for single top quark production and for Higgs boson 
production at the Tevatron and the LHC. 
From the resummation formalism we have derived NNNLO expansions 
which we use to calculate these corrections. 

For single top quark production the soft approximation works well for 
all channels at the Tevatron, and for the $s$ channel and $tW$ production 
at the LHC. The soft-gluon corrections are quite significant.

For Higgs production via the process $b{\bar b}\rightarrow H$ we also 
have to include purely collinear terms. The collinear+soft approximation is 
excellent and the corrections through NNNLO are considerable.

\mysection*{Acknowledgements}
 
This work was supported by the National Science Foundation under
Grant No. PHY 0555372.

\end{document}